\documentclass{article}
\def \inbar{\vrule height1.5ex width.4pt depth0pt}
\def \C{\relax\hbox{\kern.25em$\inbar\kern-.3em{\rm C}$}}
\def \R{\relax{\rm I\kern-.18em R}}

\newcommand{\sgn}{{\rm sgn}}

\newcommand{\id}{{\cal I}_1}
\newcommand{\idh}{{\cal I}_2}

\newcommand{\beq}{\begin{equation}}
\newcommand{\eeq}{\end{equation}}
\newcommand{\bea}{\begin{eqnarray}}
\newcommand{\eea}{\end{eqnarray}}
\newcommand{\nn}{\nonumber}

\newcommand{\pdr}{\partial}

\newcommand{\Tr}{\hbox{Tr}}
\newcommand{\Str}{\hbox{Str}}

\begin{document}
\author{Anatoly Konechny${\,}^{1}$ and  O. Teoman Turgut${\,}^{2}$\\
 \\
${}^{1}\,$Department of Physics,  University of California Berkeley  \\
and \\
Theoretical Physics Group, Mail Stop 50A-5101\\
LBNL, Berkeley, CA 94720 USA \\ 
konechny@thsrv.lbl.gov\\
\\
${}^{2}\,$Department of Physics, Bogazici University \\
80815 Bebek, Istanbul, Turkey\\ 
and\\
Feza Gursey Institute\\
 81220 Kandilli, Istanbul, Turkey\\
  turgutte@boun.edu.tr}

\title{\bf Supergrassmannian and large N limit of quantum field theory with bosons and fermions }
\maketitle
\large
\begin{abstract}
\large
We study a large $N_{c}$ limit of a two-dimensional Yang-Mills theory coupled 
to bosons and fermions in the fundamental representation. 
Extending an approach due to Rajeev we show  
 that the limiting theory can be 
described as a classical Hamiltonian system whose phase space is an 
infinite-dimensional supergrassmannian.  
The linear approximation to the equations of motion 
and the constraint  yields  the 't Hooft equations for the  mesonic spectrum. 
Two  other approximation schemes to the exact equations are 
discussed.   
\end{abstract}
\large
\section{Introduction}

To gain a better understanding of 
 gauge theories, two dimensional models are often used as a
testing ground. In a by now classic paper,  
`t Hooft has shown that  the large-$N_c$ limit allows us to 
obtain an 
equation describing the meson spectrum of two dimensional QCD 
\cite{thooft2}. The same model was analyzed using different approaches 
\cite{coote, einhorn, pak, bars} 
and they confirmed the  results obtained by `t Hooft.

In this article we  study the large-$N_c$ limit 
of certain two dimensional theories following  a general 
approach developed by  S.~Rajeev \cite{istlect,
2dqhd} (see also \cite{wadia} and
\cite{mandal} for  similar approaches). 
In the large $N_c$ limit of various quantum field  theories  (e.g., Quantum 
Chromodynamics or QCD) the quantum fluctuations become small and the theories 
are well described by  a classical limit. 
This classical limit however is different from the 
conventional one in that many of the essential non-perturbative features of 
the quantum theory survive the large $N_c$ limit\cite{thooft1, thooft2,  
witten1}. In the formulation of  \cite{2dqhd} 
the classical theory corresponding to 
large  $N_c$ limit of    2D QCD is described by a Hamiltonian system 
defined on an an infinite dimensional 
Grassmannian. The points of this infinite dimensional 
manifold   can be identified with subspaces 
in infinite-dimensional Hilbert space (see the main 
text for  precise definitions). 
The Grassmannian is a topologically nontrivial manifold 
whose connected components are labeled by an integer which
can be identified with the  baryon number.
The `t Hooft equation describing the meson mass spectrum can
 be obtained in the linear approximation
to the equations of motion on the Grassmannian  \cite{2dqhd}.  In addition to 
meson masses, this approach also allows to estimate the baryon 
mass via a variational ansatz\cite{2dbaryon, istlect}.
The overall scheme resembles the  Skyrme model of baryons in four-dimensional QCD.     
However  unlike the Skyrme model the Grassmannian system 
of  \cite{2dqhd} can be derived as a large  $N_c$ limit of 
an underlying gauge theory. 
The Grassmannian is a homogeneous manifold. It is equipped with an action 
of an infinite-dimensional group (which is unitary for the fermionic 
matter and  pseudounitary in the case of bosonic matter). 
This fact is very important for 
the structure of the phase space. In particular, it can be used 
for quantization of the classical system 
which   would allow  one to get a handle on $1/N_c$ corrections 
(including nonperturbative ones).  
 We believe that besides the possibility of describing baryons, not captured by  the original `t Hooft approach \cite{thooft1}, 
the present approach can be made  mathematically more precise. 
We remark that when the matter fields are in the adjoint representation,
the mathematical techniques required are also very elegant and interesting 
involving the Cuntz algebra in various forms. For this 
 approach, we refer the reader to 
the papers of Halpern and Schwartz \cite{halpern3} and Rajeev and Lee 
\cite{lee}.

The 2D QCD interacting with bosons in the fundamental representation
  was  also worked out following `t Hooft, partly because
bosonic theory resembles  the four dimensional QCD  in certain respects more
than the fermionic one\cite{shei, bardeen, halpern}.  The approach of \cite{2dqhd} was extended to the bosonic case 
in \cite{rajteo} (see also  \cite{tomaras} for a similar  approach to the problem).

 In this paper we  study the case when both bosonic and fermionic matter 
 are  present. One  motivation for this    
is  the fact that 
a dimensional reduction of four-dimensional QCD produces two dimensional fundamental fermions and bosons in the
adjoint representation coupled to the fermions via gauge fields. We do not
expect that the  bosons in the fundamental representation capture completely
the adjoint case, but it can be used again as a testing ground.
We also explore a more general case that includes the  Yukawa type interaction between bosons and fermions.

The  model of fundamental bosons and fermions interacting via $SU(N_{c})$ gauge field 
was studied, following the same ideas in
the original paper of `t Hooft, by Aoki\cite{aoki1, aoki2}.
The more general models in the large-$N_c$ limit are  presented in a paper by 
Cavicchi, where he uses a bilocal field approach in the 
path integral picture \cite{cavicchi}. Some of the models discussed in
\cite{cavicchi} are more general containing more   complicated
interactions, some of which  in fact require a
coupling constant renormalization.

In \cite{aoki1, aoki2, cavicchi}it is shown  that there 
are `t Hooft like spectral 
equations  for various types of mesons. In our case we have boson-boson, 
fermion-fermion,
and boson-fermion type  mesons, and they all satisfy essentially 
the same equation. In each case the meson spectrum is discrete and these
mesons are all  stable in the large-$N_c$ approximation. One cannot say much
about the  baryons using  these methods.

In the present work we  generalize the approach of \cite{2dqhd}
to 
QCD for the bosonic and fermionic matter fields coupled via gauge 
fields. We will see that the phase space of the theory corresponds to 
a certain superversion of the infinite-dimensional Grassmanian.
Although the original system does not have any supersymmetry the main 
objects describing the  large $N_c$ limit, such as the phase space, 
group action, symplectic form, 
 can be described in supergeometric terms. 
(A similar phenomenon  was observed   in another  two-dimensional  
model  in \cite{pinsky}, and indeed  this is a general feature of 
bosons and fermions coupled via gauge fields). 
We obtain the equations describing the meson spectrum 
of the model within  the linear approximation.
These equations agree with the ones   found 
by Aoki \cite{aoki1, aoki2}. The theory we will present is actually  nonlinear and can accommodate 
solitonic solutions which  should describe baryons. 
We identify the operator which gives us the baryon number.
We also propose some approximations to the spectral equations going beyond the linear approximation and
discuss some consequences.

The layout of the paper is as follows. In Section 2 we reformulate the model 
in terms of color invariant
bilinears. We further derive the Poisson brackets and the constraints imposed on the bilinear variables in large $N_{c}$ limit. 
In Section 3 we describe this Hamiltonian system in  more precise terms 
using the language of supergeometry.
 The linear approximation to the equations of motion giving the meson mass spectrum is discussed in section 4. 
In Section 5 we propose two approximation schemes that incorporate some nonlinear corrections and give 
a  qualitative discussion of their influence on the spectrum.

\section{The algebra of color invariant operators}

We start by writing  the action functionals of the two theories that we are
interested in. Both theories have a gauge field  that can be completely
 eliminated in favor of 
static 2D Coulomb potential.  We will use the light cone coordinates 
$x^{+} = \frac{1}{\sqrt{2}}( t+x)$, 
$x^{-} = \frac{1}{\sqrt{2}}( t-x)$  and choose the  $A_{+} = 0$ gauge.  
We first  look at the gauge-coupled complex bosons with
 a quartic self-interaction 
term and Dirac fermions both in the fundamental representation of $SU(N_c)$:
\begin{eqnarray}
S = && \int dx^{+}dx^{-}\bigl[-{1\over 2}\Tr F_{+-}F^{+-} +i\sqrt
2\psi_{L}^{*\alpha} (\pdr_- +igA_-)^\beta_\alpha \psi_{L\beta}+
i\sqrt 2\psi^{* \alpha}_{R}\pdr_{+}  \psi_{R \alpha} \nonumber \\
&& - m_{F}(\psi_{L}^{*\alpha}\psi_{R\alpha} + \psi_{R}^{*\alpha}\psi_{L\alpha})
  - 2\phi^{*\alpha}\pdr_-\pdr_+\phi_{\alpha}+
ig(\pdr_+\phi^{*\alpha}
{A_-}^\beta_\alpha\phi_\beta-\phi^{*\alpha}{A_-}^\beta_\alpha\pdr_+\phi_\beta) \nonumber \\
&& - m_B^2\phi^{*\alpha}\phi_\alpha - 
{\lambda^2 \over 4}\phi^{*\alpha}\phi_\alpha\phi^{*\beta}\phi_\beta\bigr] 
\end{eqnarray}

In the other model we will look at  parity broken and 
a Yukawa type interaction is added between fermions and bosons

\begin{eqnarray}
 S_{Y}=&&\int dx^+dx^-\bigl[-{1\over 2}\Tr F_{+-}F^{+-}+i\sqrt 
2\psi_L^{*\alpha}
(\pdr_- +igA_-)^\beta_\alpha\psi_{L\beta}+i\sqrt 2\psi^*_R\pdr_+\psi_R\nonumber \\
&&-2\phi^{*\alpha}\pdr_-\pdr_+\phi_{\alpha}+ig(\pdr_+\phi^{*\alpha}
{A_-}^\beta_\alpha\phi_\beta-\phi^{*\alpha}{A_-}^\beta_\alpha\pdr_+
\phi_\beta)- m_B^2\phi^{*\alpha}\phi_\alpha \nonumber \\
&&-{\lambda^2 \over 4}\phi^{*\alpha}\phi_\alpha\phi^{*\beta}\phi_\beta
+\mu(\psi_R^*\psi_{L\alpha}\phi^{*\alpha}+\psi_L^{*\alpha}\psi_R\phi_\alpha) \bigr] \, .
\end{eqnarray}
In both cases we normalize the Lie algebra generators $T^a$ as 
$\Tr T^a T^b=\delta^{ab}$, and we  choose them to be  Hermitian.
This second model is anomalous  because it is  not a chiral gauge
theory. There exist some ideas in the literature to treat an anomalous two
dimensional model \cite{jackiw}, but we will not follow this path. Instead 
we will take the above model at the classical level and eliminate 
the gauge fields which are not dynamical, and subsequently 
quantize the effective  theory.  One can check that the 
resulting system has a global $SU(N_c)$ symmetry and relativistic invariance.
We regard this as a  toy model which is {\it inspired from
gauge theory}. 

We can further use the Gauss constraint 
to eliminate the gauge field $A_{-}$ and the fermionic equations of motion 
to eliminate the right moving fermion $\psi_{R}$ ($\psi_{R\alpha}$). 
We will do these reductions in the quantized model for the first case, and 
classically for the second one.
The resulting action is first order in the ``time direction'' $x^{-}$ so we 
can pass  to Hamiltonian formalism 
in a straightforward way. 

The Fourier mode expansions read, 
$$
\phi_{\alpha}(x^{+}) = \int a_{\alpha}(p)e^{-ipx^{+}}\frac{dp}{2\pi
(2|p|)^{1/2}} \, ,  \qquad \psi_{L\alpha}(x^{+})= \int \chi_{\alpha}(p)
e^{-ipx^{+}}  \frac{dp}{2\pi 2^{1/4}},$$
(to simpify the notation   instead of
$p_+$ we  write    $p$).
The normalization factors are chosen to give the correct classical limits.
The commutation/anticommutation  relations  for the fields 
in the light cone gauge take 
the form \cite{istlect}, 
\begin{equation}
[ \chi_{\alpha}(p) , \chi^{\dagger\beta} ( q) ]_+ = 
\delta^{\beta}_{\alpha} 2\pi \delta( p - q) \, , \qquad 
[a_{\alpha}(p), a^{\dagger\beta} (q)] = {\rm sgn} (p)
\delta^{\beta}_{\alpha}2\pi\delta(p -q) \, . \end{equation}
We  define $\delta[p-q]=2\pi \delta(p-q)$, and use $[dp]={dp\over 2\pi}$
to keep track of factors of $2\pi$.

One defines a   Fock vacuum state $|0\rangle $ by  the conditions,
\beq
 a_{\alpha}(p)|0\rangle = \chi_{\alpha}(p)|0\rangle =0\ \ {\rm for }\ \ p>
0 \quad   a^{\dagger\alpha}(p)|0\rangle =
 \chi^{\dagger\alpha}(p)|0\rangle =0 \  \  {\rm for}\ \  p< 0.
\eeq 
The corresponding normal orderings are defined as
\begin{equation}  
: \chi^{\dagger\alpha}(p) \chi_{\beta}(q) : = 
\left\{ \begin{array}{ll} 
-\chi_{\beta} (q) \chi^{\dagger\alpha}(p) & \mbox{if}\enspace  p<0, q<0, \\
\chi^{\dagger\alpha}(p) \chi_{\beta}(q) & \mbox{otherwise} \, ,\\
\end{array} \right. 
\end{equation} 
\begin{equation}  
: a^{\dagger\alpha}(p) a_{\beta}(q) : = 
\left\{ \begin{array}{ll} 
a_{\beta} (q) a^{\dagger\alpha}(p) & \mbox{if}\enspace  p<0 , q<0, \\
a^{\dagger\alpha}(p) a_{\beta}(q) & \mbox{otherwise} \, .\\
\end{array} \right. 
\end{equation} 
(Later on we also use the extension of normal ordering  to product of four 
operators, and it is the standard one).
For our purposes it is most convenient to remember the normal 
orderings of bilinears in the following form:
\bea
  :\chi^{\alpha\dag}(p)\chi_\beta(q):&=&\chi^{\alpha\dag}(p)\chi_\beta(q)
-{\delta^\alpha_\beta\over 2}[1-\sgn(p)]\delta[p-q]\nn\cr
:a^{\alpha\dag}(p)a_\beta(q):&=&a^{\alpha\dag}(p)a_\beta(q)-
{\delta^\alpha_\beta\over 2}[1-\sgn(p)]\delta[p-q].
\eea
Written as quantum operators, we have in the first model,
\beq
   \psi_{R\alpha}={m_F \over \sqrt{2}i\pdr_+}\psi_{L\alpha}
\eeq
and its hermitian conjugate, and 
\beq
 \psi_R=-{\mu \over \sqrt{2}i\pdr_+}\phi^{*\alpha}\psi_{L\alpha}
\eeq
and its hermitian conjugate for the second model.
In the first case, $A_-$ is given in terms of the other fields as,
\beq
    A_-^a=-{g \over \pdr_+^2}:(\sqrt{2}\psi^{\dagger\alpha}_L
(T^a)^{\beta}_{\alpha} 
\psi_{L\beta}+i[\phi^{\dagger\alpha}(T^a)^{\beta}_{\alpha}\pdr_+\phi_\beta
-\pdr_+\phi^{\dagger\alpha}(T^a)^{\beta}_{\alpha}\phi_\beta]): \eeq
In the second model we are using the same equation to 
eliminate $A_-$ at the classical level
(which  means without the normal ordering).

By eliminating the redundant degrees of freedom we can express 
the action in terms of the 
bilinears of the fields $\psi_{L\alpha}$ and $\phi_\alpha$ only.
We introduce,  
\bea
\label{bilin}  \hat  M(p,q)&=&{2 \over  N_c}
:\chi^{\dagger \alpha}(p)\chi_\alpha(q):\cr 
  \hat  N(p,q)&=&{2 \over  N_c}:a^{\dagger \alpha}(p)a_\alpha(q):
\eea
and their odd counterparts, 
\bea 
 \hat Q(p,q)={2\over N_c}\chi^{\dagger\alpha}(p)a_\alpha(q) \, ,
\quad \hat{\bar Q}(r,s)={2 \over N_c}a^{\dagger \alpha}(r)\chi_\alpha(s)
\eea

Once the redundancies are removed the 
resulting action is already first order in the ``time'' variable hence we can
read off the Hamiltonian, and the resulting commutation relations are
consistent with the ones obtained from the conventional canonical 
quantization.
The reduction is straightforward in principle but requires a long and 
careful computation. Since the details are explained in Rajeev's lecture notes
\cite{istlect}
we only give the result:
\beq
         H=H_0+H_I,
\eeq
\beq
    H_0={1\over 4}M_B^2\int {[dp]\over |p|}N(p,p)+
{1\over 4}M_F^2\int {[dp]\over p}M(p,p),
\eeq
where 
for the first model we use,
\beq 
  M_F^2=m_F^2-{g^2\over \pi},\quad M_B^2=m_{RB}^2-{g^2\over \pi}
.\eeq
We employ  a logarithmic renormalization on the bare mass of the bosonic
field \cite{tomaras}, and denote the renormalized mass 
as $m_{RB}$. 
For the second model, since we reduce it at the classical level 
there are no corrections coming to the boson  mass term,
\beq
        M_F^2=0, \quad M_B^2={m_B^2\over 4}
.\eeq

The interaction parts are given by
\bea 
H_I&=&\int [dpdqdsdt]\ G_1(p,q;s,t)M(p,q)M(s,t)+\int [dpdqdsdt]\ G_2(p,q;s,t) 
N(p,q)N(s,t)\nn\cr 
&+&\int [dpdqdsdt]\  G_3(p,q;s,t)Q(p,q)\bar Q(s,t) 
,\eea 
where both for the first and second models,
\beq  
\label{hamil}
G_1(p,q;s,t)=-{g^2\over 16}\Big({1 \over (p-t)^2}+{1 \over
(q-s)^2}\Big)\delta[p+s-t-q]  \eeq 
\beq 
\label{hamilt}
G_2(p,q;s,t)={g^2\over 64}\Big({1 \over (p-t)^2}+{1 \over (q-s)^2}\Big) 
\delta[p+s-t-q] {qt+ps+st+pq \over \sqrt{|pqst|}} 
+{\lambda^2\over 64}{\delta[p+s-t-q] \over \sqrt{|pqst|}} 
.\eeq 
In the first model we use,
\beq 
\label{hamilto}
G_3(p,q;s,t)={g^2\over 8}{q+s\over (q-s)^2}{\delta[p+s-t-q]\over
\sqrt{|qs|}}  ,\eeq 
and for the second model we only have an additional term,
\beq 
\label{hamilton}
G_3(p,q;s,t)={\mu^2\over 16}{1\over (p-q)}{1 \over \sqrt{|qs|}} 
\delta[p-t-q+s] 
+{g^2\over 8} {q+s\over (q-s)^2}{\delta[p+s-t-q]\over \sqrt{|qs|}} 
.\eeq
Above we rescaled our coupling constants by a factor of $N_c$ and
keep the same symbols  for the couplings(so $g^2N_c\mapsto g^2$,
$\mu^2N_c\mapsto \mu^2$ and $\lambda^2N_c\mapsto \lambda^2$)
to simplify notation.  
For the precise meaning of these singular integral kernels 
we refer to the lecture notes of  Rajeev\cite{istlect}:
we should interpret them as Hadamard principal value. We will continue to
write the ordinary integrals but keep in mind that the integrals are
evaluated with this prescription. 

The theory we obtained still possesses a global $SU(N_{c})$ invariance.
 The corresponding generator of symmetry   is 
\begin{eqnarray} \label{color}
\hat {\cal Q}^{\alpha}_{\beta} &=& \int [dp] \left( :\chi^{\dagger\alpha}(p) 
\chi_{\beta}(p) :
 - \frac{1}{N_{c}} \delta^{\alpha}_{\beta} 
: \chi^{\dagger\gamma}(p) \chi_{\gamma}(p) :\right) + \nonumber \\ 
&+& \int [dp] \sgn(p)\left( : a^{\dagger\alpha}(p) a_{\beta}(p) :
 - \frac{1}{N_{c}}  \delta^{\alpha}_{\beta} 
 : a^{\dagger\gamma}(p) a_{\gamma}(p) : \right) \, .   
\end{eqnarray}
It is known (at least for the purely spinor and purely  scalar $\rm QCD_{2}$)
 that in the light-like axial gauge only the color singlet sector 
of the model can be quantized in a way that preserves Lorentz invariance
  (\cite{bars,pak}). 
In this paper we will therefore consider only the restrictions 
 of our models to this  sector.  
In general  for a gauge theory it  is expected that 
in the large $N_c$ limit any gauge invariant  correlator factorizes, i.e. 
$<AB> = <A> <B> + O(1/{N_c}) $. So when  the two dimensional  
theory restricted to the color invariant subspace in the large $N_c$ limit any 
color invariant correlator  should be 
expressible in terms of correlators of color invariant bilinear operators,
$\hat M$, $\hat N$ and $\hat Q, \hat{\bar Q}$ given in (\ref{bilin}) and (11).
We  compute
 the (anti)commutation relations between these bilinears:

\begin{eqnarray} \label{ccr}
[ \hat M(p,q),\hat M(r,s)]&=& \frac{2}{ N_{c}} \bigl[ \hat M(p,s)\delta[q-r] -
 \hat M(r,q)\delta[p-s]\nn\\ 
 &-&  \delta[p-s]\delta[q-r](\sgn(p)-\sgn(q)) \bigr] \,  \nn\\ 
 \left[ \hat N(p,q),\hat N(r,s) \right] & =&  
\frac{2}{N_{c}} \bigl[ \hat N(p,s)\sgn(q)\delta[q-r]-\hat N(r,q)\nonumber
\sgn(p)\delta[p-s] \nn\\
&+ & \delta[q-r]\delta[p-s](\sgn(p)-\sgn(q)) \bigr] \nn\\ 
\left[ \hat Q(p,q),\hat{\bar Q}(r,s)\right]_+ &=&\frac{2}{N_{c}}
\bigl[ \hat M(p,s)\sgn(q)\delta[q-r]+\hat N(r,q)
\delta[p-s]\nn \\
&+&\delta[p-s]\delta[q-r](1-\sgn(p)\sgn(q)) \bigr] \,  \nn \\
\left[ \hat M(p,q),\hat Q(r,s)\right]&=&\frac{2}{N_{c}}\delta[q-r] \hat Q(p,s)
\nn \\
\left[ \hat N(p,q),\hat Q(r,s)\right]&=&-\frac{2}{N_{c}}\delta[p-s]\sgn(p) 
\hat Q(r,q)\nn \\
\left[ \hat M(p,q),\hat{\bar Q}(r,s)\right]&=&-\frac{2}{N_{c}}\delta[p-s]
\hat{\bar  Q}(r,q)\nn \\
\left[ \hat N(p,q),\hat{\bar Q}(r,s)\right] &=&\frac{2}{N_{c}}\delta[q-r]
\sgn(q)\hat{\bar Q}(p,s)
\eea

All the other (anti)commutators  vanish.
The first two relations were used before  in \cite{2dqhd} and 
\cite{rajteo} respectively. 
These (anti)commutation relations define an infinite dimensional  Lie 
superalgebra. 
Its even part is isomorphic 
to a direct sum of central extensions of infinite-dimensional unitary
 and pseudo unitary groups each one generated by operators 
$\hat M(p,q)$ and $\hat N(p,q)$ respectively (see \cite{rajteo} for details).
 We will talk more about this Lie superalgebra and the corresponding 
supergroup in the next section.  
As the right hand sides of (\ref{ccr}) all contain 
a factor of $1/N_{c}$ in the 
large $N_{c}$ limit all of the bilinears 
commute (or anticommute respectively) and can be thought of 
as coordinates on a classical phase space. 
We denote the classical variables corresponding to 
$\hat M$, $\hat N$, $\hat Q$, $\hat{\bar Q}$ by the same letters with hats
 removed. 
This classical phase space is an infinite dimensional 
supermanifold endowed with a super Poisson structure inherited from 
the (anti)commutation relations (\ref{ccr}). The corresponding 
Poisson superbrackets are obtained from the (anti)commutators 
in (\ref{ccr}) by substituting $-i$ instead of     $1/N_{c}$  factors
(note that this brings an extra factor of $2$). 
There is no simple way to decide 
which multiple of  $1/N_c$ should be 
 the quantum parameter.  
If one attempts a  geometric quantization of
 this model, the symplectic form should be an   
integer multiple of the Chern character of the line bundle, 
the sympectic form we have in the next 
section is in fact the basic two form. The other possibility is to write 
the symplectic form in
the action and use single valuedness of the path integral as is done 
in \cite{2dqhd}. 
(We note in passing that there is a 
factor of $2$ missing in the reference \cite{rajteo}, due to an error 
in the conventions, but we scale the 
Hamiltonian with the same parameter so the 
large-$N_c$ results are the same. The geometric quantization parameter
instead should have  been  $1/N_c$).

However the super-Poisson structure
corresponding to (\ref{ccr}) only gives  a local structure of the classical
phase space  of the theory. In addition to that there are some global
constraints  on the classical variables assigned to the color  invariant
bilinears. The constraints emerge in the large $N_{c}$ limit  as consequences
of  the color invariance  condition $\hat {\cal Q}^{\alpha}_{\beta} = 0$.

To write down these constraints it is convenient to introduce the 
following operator product convention
$$
(AB) (p,q) = \int[ dr]\, A(p, r) B(r, q) 
$$
where $A$, $B$ stand for any of the above (classical) bilinears.  
We also introduce operators 1 and $\epsilon$ as the operators with the 
 kernels 
$\delta[p-q]$ and $-{\rm sgn}(p)\delta[p-q]$, respectively.
In this notation the constraints read as follows
\begin{eqnarray} \label{constr}
    (M+\epsilon)^2+Q\epsilon  Q^\dag &=&1\nn \\
     \epsilon  Q^\dag M+\epsilon  Q^\dag  \epsilon +\epsilon N 
\epsilon Q^\dag + Q^\dag &=&0\nn \\
     M Q+\epsilon Q+Q\epsilon N+Q\epsilon&=&0\nn \\
 (\epsilon N+\epsilon)^2+\epsilon Q^\dag Q&=&1\, . 
\end{eqnarray}

For brevity we will present here  a derivation only of  
 the first constraint in (\ref{constr}). The derivations of all 
the others are very similar. 
We will restrict ourselves to  the zero subspace of the operator  
$\hat {\cal Q}^{\alpha}_{\beta}$ and  we   define the number operators
\begin{equation} \label{part} \hat F \equiv \frac{1}{N_{c}}\int [dp]
:\chi^{\dagger\alpha}(p) \chi_{\alpha}(p):  \, , \quad 
\hat B \equiv \frac{1}{N_{c}}\int [dp] \sgn(p) :a^{\dagger\alpha}(p)
a_{\alpha}(p): \quad  
.\end{equation}  
(Note that these operators are scaled by a factor of $\frac{1}{N_{c}}$ so
taking  the limit $N_{c}\to \infty$ 
gives us zero when these operators are acting on mesonic states.
They are nonzero when we look  at the baryonic states as we will see
shortly.) 

By writing out the product of operators at hand in terms of the variables $a$, $a^{\dagger}$, 
$\chi$ and $\chi^{\dagger}$ and moving the suitable  combinations 
to the right  one can prove the identity (that holds on the whole Fock  space)   
$$
((\hat M + \epsilon )^{2} + \hat Q\epsilon \hat Q^{\dagger})(r,s)  =
\delta[r-s] + \frac{2}{N_{c}^{2}}\chi^{\dagger\alpha}(r) \chi_{\beta}(s) (
\hat {\cal Q}_{\alpha}^{\beta}  + 
\delta^{\beta}_{\alpha} (\hat F + \hat B)) $$ 
On the subspace $\hat {\cal Q}_{\alpha}^{\beta}=0$,  the operator 
$\hat B+\hat F$ will be equal to  the baryon
number operator  $\hat{\bf B}$. 
Thus when we restrict  ourselves to a fixed baryon 
number $B$, we get,

$$   
((\hat M + \epsilon )^{2} + \hat Q\epsilon \hat Q^{\dagger})(r,s) =
\delta[r-s] + (\hat M + 1 - \epsilon) (r,s)\frac{B}{N_c} \, ,  $$
this  in the large $N_{c}$ limit produces the first constraint in
(\ref{constr}).

When we look 
at a possible exotic baryon state:
\beq
       \int\epsilon_{\alpha_1...\alpha_s\alpha_{s+1}...\alpha_{N_c}}
 Z(p_1,...,p_s;p_{s+1},...,p_{N_c})
\chi^{\dagger\alpha_1}(p_1)...\chi^{\dagger\alpha_s}(p_s)
a^{\dagger\alpha_{s+1}}(p_{s+1})...a^{\dagger\alpha_{N_c}}(p_{N_c})|0>
,\eeq
where $p_1,p_2,...p_{N_c}$ are all positive, 
and $Z(p_1,...,p_s,p_{s+1},...p_{N_c})$ is symmetric in
$p_1,...,p_s$ and  antisymmetric in $p_{s+1},...,p_{N_c}$. The operator
$\hat {\bf B}$ gives $1$ acting on such 
a state. On mesonic states this operator has vanishing matrix
elements in the large-$N_c$ limit.
One can prove more generally therefore that this operator is the baryon
number operator. If we act by this operator on a product of such exotic
baryons and finite number of mesons in the large-$N_c$ limit we get the number
of baryons, $B$. In this discussion we see the possibility of having exotic
baryons, and we will come back to the geometric meaning of this in the next
section (and show that it is indeed an integer in our model). 
We will also show that 
 just as in the purely bosonic and purely fermionic cases the
constraints  (\ref{constr}) have 
an elegant geometric interpretation in terms of infinite dimensional  disc and Grassmannian.

\section{Phase Space of the Theory: Super-Grassmannian}

In this section we  present  the 
geometry of the phase space without going into the mathematical intricacies. 
We believe  the most proper treatment requires an infinite dimensional 
extension of Berezin's ${\bf Z}$-graded version of super-geometry.
We do not give such a complete presentation, and develop  a more 
formal approach (in many cases we provide paranthetical remarks
on the general case).
We plan to provide  a more detailed discussion  in a later publication when we
discuss geometric quantization of this system. The proper treatment of
second quantization with bosons and fermions, which fits to our
point of view, can be found in \cite{langmann1, langmann2},
and also  in \cite{langmann3}.
In order to understand the geometry of the phase space, we  define an operator 
in  super-matrix form;
\beq
   \Phi= \pmatrix{\epsilon N +\epsilon& \epsilon Q^\dag \cr 
Q  &  M+\epsilon}
,\eeq
where $\Phi:{\cal H}^e|{\cal H}^o\to {\cal H}^e|{\cal H}^o$.
We think of the direct sum ${\cal H}^e \oplus {\cal H}^o $
 of one-particle Hilbert spaces
of bosons and fermions respectively  
 as even and odd graded
and the notation ${\cal H}^e|{\cal H}^o$ is used to emphasize this grading.
We use $\epsilon=\pmatrix{-1&0\cr 0&1}$ in both 
of these spaces. This matrix realization corresponds to 
the decomposition of the Hilbert spaces into positive and 
negative energy subspaces as ${\cal H}_+^e\oplus {\cal H}_-^e$ for 
bosons and ${\cal H}_+^o\oplus {\cal H}_-^o$ for fermions.

The  constraints and the conditions that we found 
in the previous section 
on the basic variables of our theory in terms of $\Phi$  become 
\beq
  \Phi^2=1 \quad E\Phi^\dag E=\Phi
,\eeq
where $E=\pmatrix{\epsilon &0\cr 0& 1}$.

If we introduce a super-group of operators 
acting on ${\cal H}^e|{\cal H}^o$, obeying the relations  
\beq
    gEg^\dag=E, \quad g^\dag E g=E,
\eeq
we see that the  action of this group on the variable $\Phi$,
 $(g,\Phi)\mapsto g\Phi g^{-1}$ preserves the above stated conditions on 
$\Phi$.
The orbit of $\hat  \epsilon=\pmatrix{\epsilon &0\cr 0 &\epsilon}$
under the action of this super-unitary group,
can be parametrized by $\Phi$.

The condition that the  bilinears, originally 
defined on the  Fock space of the quantum theory, create finite norm 
vectors implies that the off-diagonal components of
$M$ and $N$ are Hilbert-Schmidt operators(see \cite{2dqhd, istlect,
rajteo} and for the ideals in the non-super case see \cite{gohberg, simon}).
A similar computation shows that 
the off-diagonal components of the super-operators 
$Q$ and $Q^\dag$ also satisfy these conditions.
These finite norm conditions in two dimensions can 
be written in an economical way as the   Hilbert-Schmidt 
condition on the super-matrix $\Phi$. 
To state these convergence conditions more properly we should decompose
${\cal H}^e|{\cal H}^o$ into negative and positive energy subspaces and
think  
of $\Phi$ as an operator acting from 
${\cal H}^e_+|{\cal H}^o_+\oplus {\cal H}^e_-| {\cal H}^o_-$ to the same 
space.
Thus we have   the  convergence conditions, 
\beq 
   [ \hat \epsilon, \Phi] \in {\cal I}_2 \, \quad \hat \epsilon=
\pmatrix{-1&0\cr 0&1}
,\eeq
where we write $\hat \epsilon$'s matrix realization  with respect to
this positive-negative energy decomposition.
Here  for the upper off-diagonal 
component the ideal of Hilbert-Schmidt operators ${\cal I}_2$ refers 
to the set of operators  
$B:{\cal H}^e_-|{\cal H}^o_-\to {\cal H}^e_+|{\cal H}^o_+$,
such that $\Tr B^\dag B$ 
is convergent. (This definition in the 
${\bf Z}_2$ graded case is  the usual one, since the operators have  ordinary
numbers as their matrix entries, in a fully ${\bf Z}$ graded case these 
questions are delicate and we have to give a precise meaning to the 
Hilbert-Schmidt condition. For this work we ignore this question but 
see \cite{superdisk} for further comments on it).
The lower off-diagonal block will be a Hilbert-Schmidt 
operator 
$C:{\cal H}^e_+|{\cal H}^o_+\to {\cal H}^e_-|{\cal H}^o_-$ as well.
(Since the even and odd Hilbert spaces are isomorphic, it is 
convenient to drop the superscript when there is no confusion).  
The above considerations suggest  
that we should use as our symmetry group the  restricted super-unitary group:
\beq
   U_1({\cal H}_-,{\cal H}_+|{\cal H})=\Big\{g|\  gEg^\dag=E, \quad 
g^\dag E g=E \ \quad 
[\hat \epsilon ,g]  \in {\cal I}_2\Big\}
,\eeq
where ${\cal I}_2$ denotes the ideal of Hilbert-Schmidt 
operators as in the above positive-negative energy decomposition
used for $\Phi$'s convergence conditions.

We look at the orbit of $\hat \epsilon=\pmatrix{\epsilon & 0\cr 0 & \epsilon}$,
this time we write it with respect to the original 
decomposition ${\cal H}^e|{\cal H}^o$, 
 under the  restricted super-unitary group.
We notice that this  orbit is in fact 
a homogeneous super-symplectic manifold:
\beq
    SGr_1={ U_1({\cal H}_-,{\cal H}_+|{\cal H} )\over U({\cal H}_-|{\cal H}_-)\times 
   U({\cal H}_+|{\cal H}_+)}
.\eeq
The stability subgroup has a natural embedding into the full group.
This physically means that we allow mixing of the  
positive energy states of bosons and fermions as well as the negative ones.

Notice that a tangent vector $V_u$ at any point on this 
super-Grassmannian is 
given by its effect on $\Phi$,
$V_u(\Phi)=i[u,\Phi]_s$, where 
we use the super-Lie bracket which is defined by
\bea
    \Big[ \pmatrix{a_1&\beta_1\cr \gamma_1&d_1},
\pmatrix{a_2&\beta_2\cr \gamma_2&d_2}
\Big]_s&=&\Big[\pmatrix{a_1&0\cr 0&d_1},\pmatrix{a_2&0\cr 0&d_2}\Big]+
    \Big [\pmatrix{a_1& 0\cr 0 &d_1},\pmatrix{0&\beta_2\cr \gamma_2& 0}\Big]\cr\nn
&+&\Big[\pmatrix{0&\beta_1\cr \gamma_1&0},\pmatrix{a_2&0\cr 0&d_2}\Big]+
\Big[\pmatrix{0&\beta_1\cr \gamma_1&0},\pmatrix{0&\beta_2\cr \gamma_2& 0}\Big]_+
\eea
for a decomposition of $u$ into 
$\pmatrix{a&\beta\cr \gamma& d}$, with respect to ${\cal H}^e|{\cal H}^o$.
In general the super-Lie algebra element $u$ will 
depend on the position $\Phi$.

Our  homogeneous manifold carries a  natural two-form, this 
turns it into a  phase 
space.  We  formally define a   two-form;
\beq
  \Omega={ i \over 4} \Str\Phi d\Phi \wedge d\Phi
.\eeq
One can give the symplectic form 
explicitly via its action on vector fields, and this defines the 
above two form:
\beq
  i_{V_u}i_{V_v}\Omega={i \over 8}\Str \Phi [[u,\Phi]_s,[v,\Phi]_s]_s
.\eeq
Using exactly the   
same methods as in \cite{2dqhd,rajteo}, we can show that it is 
closed and non-degenerate.

In fact the above form  is also  a  homogeneous two-form
invariant under the group action, as can be verified in a simple manner.
We note  that the  super-Poisson 
brackets which we introduced in the first 
section as a result of the large-$N_c$ limit, 
are precisely the ones given by this symplectic form.
Therefore we may introduce a classical dynamical system 
defined on this super-Grassmannian with this symplectic
form which gives us the same  set of 
super-Poisson brackets. 
This shows that the large-$N_c$ limit of our theory 
has an independent geometric formulation: the phase space is 
an infinite dimensional homogeneous manifold with a  natural
symplectic structure on it.

The group action is generated by moment maps 
$F_u=-{i\over 2}\Str_\epsilon u(\Phi-\pmatrix{\epsilon&0\cr 0&\epsilon})$,
where we use the even-odd decomposition to write 
all the operators and conditional trace to be defined below.  
They   satisfy the
following super-Poisson realization of the super-unitary group:
\beq
 \{ F_u,F_v\}=F_{-i[u,v]_s}-
{i\over 2}\Str_\epsilon\Big[\pmatrix{\epsilon&0\cr 0&\epsilon},u\Big]_s v
,\eeq
where $[.,.]_s$ again  denotes the super-commutator(super-Lie bracket).
To see this, one way is to compute both sides, the other is to use general 
principles and evaluate both sides at $\hat \epsilon=\pmatrix{\epsilon&0\cr
0&\epsilon}$ (written with respect to the even-odd decomposition).
The moment function on the right vanishes there and the central term 
is constant on the phase space,
this gives us, 
\bea
    \Sigma_s(u,v)&=&-{i\over 8} \Str
 \pmatrix{\epsilon&0\cr 0&\epsilon}
\Big[\Big[\pmatrix{\epsilon&0\cr 0&\epsilon},u\Big],
  \Big[\pmatrix{\epsilon&0\cr 0&\epsilon}, v\Big]\Big]_s\cr\nn
   &=& -{i\over 2}\Big(\Tr_{\epsilon}[\epsilon, a(u)]a(v)
-\Tr_\epsilon([\epsilon, \beta(u)]\gamma(v)+[\epsilon,\beta(v)]\gamma(u))
-\Tr_{\epsilon}[\epsilon, d(u)]d(v)
\Big)\nn\cr
  &=&-{i \over 2} \Str_\epsilon 
\big[\pmatrix{\epsilon& 0\cr 0&\epsilon}, u\big]v
.\eea

The conditional super-trace is defined by 
$\Str_\epsilon \pmatrix{A&B\cr C&D}=\Tr_\epsilon A-\Tr_\epsilon D$, and 
$\Tr_\epsilon A={1\over 2}\Tr (A+\epsilon A\epsilon)$.
Notice that the convergence conditions on $\Phi$   guarantees that the 
conditional trace exists (in fact a better convergence is possible, see 
below).
This can be seen most easily by using,
$\Phi-\hat \epsilon=g\hat  \epsilon g^{-1}-\hat \epsilon
=-[\hat \epsilon,g]g^{-1}$. It is more natural to compute this 
with respect to the positive-negative energy decomposition,
(we use the subscripts $\pm$ to denote the super-matrix 
elements acting between various subspaces), 
\beq
    [\hat \epsilon, g]g^{-1}=\pmatrix{0& g_{+-}\cr g_{-+}&0}
\pmatrix{(g^{-1})_{++}&(g^{-1})_{+-}\cr (g^{-1})_{-+}&(g^{-1})_{--}}
=\pmatrix{\id&\idh\cr\idh&\id}
,\eeq
where ${\cal I}_1$ denotes the ideal of  trace class operators and 
${\cal I}_2$ is the ideal of Hilbert-Schmidt operators.
We used the fact that the off diagonal elements are Hilbert-Schmidt and
the others are bounded, and the analog of the well-known conditions 
${\cal I}_2{\cal I}_2\in {\cal I}_1$ in the super-case.
If we multiply this with an element of the Lie algebra we see that the 
conditional traces exist.
It suggests a slightly better way to write the moment maps,
$F_u=-{i\over 2}\Str_{\hat \epsilon} u(\Phi-\hat\epsilon)$, which 
shows that the conditional convergence could be actually
defined with respect to the positive-negative energy decomposition.

The above discussion  further implies that 
$\Str_{\hat \epsilon} (\Phi-\hat  \epsilon)$ is convergent. 
This expression is in
fact conserved by the equations of motion of  a quadratic Hamiltonian.
We may understand the meaning of this number, if we think of its action on
color invariant states  before we take the large-$N_c$ limit.
We can prove that in this case this operator gives us twice the  baryon 
number. Recall that the baryons in this theory can be exotic, that
is   we may have color singlet combinations of the form,  
\beq
\int\epsilon_{\alpha_1\alpha_2...\alpha_{N_c}}
 Z(q_1,...q_s;q_{s+1},...,q_{N_c})
\chi^{\dagger\alpha_1}(q_1)... 
\chi^{\dagger\alpha_k}(q_k)a^{\dagger\alpha_{k+1}}(q_{k+1})
...a^{\dagger\alpha_{Nc}}(q_{N_c})|0>
,\eeq
where all the momenta are positive, and $Z$ is symmetric in
$p_1,...,p_s$ and antisymmetric in $p_{s+1},...,p_{N_c}$,
as we have seen in the previous section. The negative momenta case,
\beq
\int \epsilon^{\alpha_1\alpha_2...\alpha_{N_c}}
{\bar Z}(q_1,...q_k; q_{k+1},...,q_{N_c})\chi_{\alpha_1}(q_1)... 
\chi_{\alpha_k}(q_k)a_{\alpha_{k+1}}(q_{k+1})...a_{\alpha_{Nc}}(q_{N_c})|0>
,\eeq
where all the momenta negative, and 
similar symmetry properties for $\bar Z$ corresponds to 
an anti-baryon and $\hat{\bf B}$
acting on such a state 
gives $-1$.  
 So we identify  the
large-$N_c$ limit of the baryon number operator as, 
\beq
     {\bf B}=-{1\over 2} \Str_{\hat\epsilon} (\Phi-\hat \epsilon)
.\eeq

We show in the appendix A  that the baryon number operator is indeed 
an integer using the geometry of our phase space.
We will leave the discussion of the geometry of the phase space at this point 
and return to the dynamics.

\section{The Linear Approximation}

In this section we discuss the linear approximation to the above theory. At
present we  do not have a simple physical  interpretation of
the full equations of motion.
In principle they are 
straightforward to compute using the Hamiltonians we have and the 
defining Poisson brackets.
Our phase phase is  defined by the Poisson brackets we get  
from the super-commutators for this system in the large-$N_c$ 
limit and the constraints which define the 
global nature of the phase space. We note that part of the 
interactions of this theory are in these constraints.
We give the super Poisson brackets, that  defines the kinematics of our theory: 
\begin{eqnarray}  
 \{ M(p,q), M(r,s)\}&=&- 2i \bigl[  M(p,s)\delta[q-r] -  
  M(r,q)\delta[p-s]\nn\\ 
  &-&  \delta[p-s]\delta[q-r](\sgn(p)-\sgn(q)) \bigr] \,  \nn\\ 
   \{  N(p,q), N(r,s) \} & =&   
- 2i \bigl[  N(p,s)\sgn(q)\delta[q-r]- N(r,q)\nonumber  
\sgn(p)\delta[p-s] \nn\\
&+ & \delta[q-r]\delta[p-s](\sgn(p)-\sgn(q)) \bigr] \nn\\
 \{  Q(p,q), {\bar Q}(r,s) \}_+  
&=&-2i  \bigl[  M(p,s)\sgn(q)\delta[q-r]+ N(r,q) \delta[p-s]\nn \\  
&+&\delta[p-s]\delta[q-r](1-\sgn(p)\sgn(q)) \bigr] \,  
 \nn \cr \{  M(p,q), Q(r,s)\} 
&=&-2i\delta[q-r] Q(p,s) \nn \\
\{ N(p,q), Q(r,s)\}&=&2i\delta[p-s]\sgn(p)   
 Q(r,q)\nn \cr \{  M(p,q),{\bar Q}(r,s)\} 
&=&2i\delta[p-s] {\bar  Q}(r,q)\nn \\  
\{ N(p,q),{\bar Q}(r,s)\} &=&-2i\delta[q-r]  
\sgn(q){\bar Q}(p,s)  
.\end{eqnarray}  
We have   the constraints for the basic variables given in 
equation (21).

If we are given a Hamiltonian we can compute the equations of
motion using the above super-Poisson brackets. This is a complete 
description of a classical system. Of course since the 
theory is infinite dimensional there are various delicate 
questions, such as, is it possible to define trajectories for a 
any given initial data, what is the dense domain on which the Hamiltonian 
is defined, etc. We will not attempt 
to answer these quaestions here. 
In the limit $N_c \to \infty$, we can rewrite  the Hamiltonians of 
interest  
in terms of these classical variables, the answers are given in  
Section 2. 
\beq 
   H=H_0+H_I 
,\eeq 
here $H_0=\int [dp] h_F(p)M(p,p)+\int [dp] h_B(p) N(p,p)$, 
and we take $h_F(p)={M_F^2\over 4}{1\over p}$ and 
$h_B(p)={M_B^2\over 4}{1\over |p|}$ with the interpretation that 
these mass terms are given by the previous expressions.
$H_I$, the interaction part, is given generally by  
\bea 
H_I&=&\int [dpdqdsdt]\ G_1(p,q;s,t)M(p,q)M(s,t)+\int [dpdqdsdt]\ G_2(p,q;s,t) 
N(p,q)N(s,t)\nn\cr 
&+&\int [dpdqdsdt]\  G_3(p,q;s,t)Q(p,q)\bar Q(s,t) 
.\eea 
In the next section, it will be useful to keep this
 general form of the Hamiltonian, 
but their explicit forms are given in the discussion of the models in 
Section 2  in (\ref{hamil}),(\ref{hamilt}), (\ref{hamilton}),
 we will use them directly 
(in the calculations we keep $\mu^2$ always, for the first model we 
must  set $\mu^2=0$).

It is straightforward to find the resulting non-linear equations of  
motion simply by computing 
\beq 
   { \partial O(x^-) \over \partial x^-}=\{ O(x^-), H\}_s, 
\eeq 
for any observable $O$ of the theory (we allow for 
an odd Hamiltonian in the above form, but in our cases,
the Hamiltonians are even). 
 However, it is simpler to first look at the  linearization where everything
decouples (equations for   $M$ and $N$ were analyzed in this approximation in
previous publications  \cite{2dqhd, rajteo, istlect}). We will see that 
we  get the same equations for $M, N$ as in \cite{2dqhd, rajteo}
 in our  linearized theory.

Let us ignore all the quadratic terms in the equations of motion and 
all the quadratic terms in the constraints. First let us write down the 
resulting constraints in this approximation:
\bea
      &\ &\epsilon M+M\epsilon=0\nn\cr
      &\  &\epsilon N\epsilon+N=0\nn\cr
      &\  & \epsilon Q^\dag \epsilon+Q^\dag=0\nn\cr
      &\  & \epsilon Q+Q\epsilon=0
.\eea
We note that the last two equations are identical and the constraints on these
variables decouple hence they can be solved independently.
The solutions are,
\beq
     M(u,v)=0, \quad N(u,v)=0, \quad Q(u,v)=0\quad {\rm for} \quad uv>0
.\eeq
The other components, that is the ones which have 
opposite sign momenta,  are not restricted.
The equations of motion one gets for the variable $M$ in the linear
approximation is 
 (for $u>0,v<0$):
\beq
   {\partial M(u,v;x^-)\over \partial x^-}=i{M_F^2\over 2}\Big({1\over u}
-{1\over v}\Big)M(u,v)-{ig^2\over 2\pi}\int_{-{u-v\over 2}}^{u-v\over 2}
dp{M(p-{u-v\over 2},p+{u-v\over 2})\over (p-{u+v\over 2})^2}
,\eeq
which is identical with the one in \cite{2dqhd,istlect}.
If we make the ansatz (see \cite{2dqhd}) $M(u,v)=\xi_M(x)e^{iP_-x^-}$,
where $x={u\over u-v}$, and define  the invariant mass
$\Lambda_M^2=2P_-(u-v)$,(recall that $(u-v)=P_+$), we get,
\beq
   \Lambda_M^2\xi_M(x)=M_F^2\Big({1\over x}+{1\over 1-x}\Big)\xi_M(x)
-{g^2\over \pi}
\int_0^1 {dy\over (y-x)^2}\xi_M(y)
.\eeq 
This is the well-known `t Hooft equation \cite{thooft1}.
Similarly for $N(u,v)$ using the same type of ansatz,
$N(u,v)=\xi_N(x)e^{iP_-x^-}$, and the invariant mass
$\Lambda_N^2=2P_-(u-v)$, we get,
\begin{eqnarray}
\Lambda_N^2\xi_N(x)&=&M_B^2\Big({1\over x}+{1\over 1-x}\Big)\xi_N(x)
-{g^2\over 4\pi}\int_0^1{dy\over (y-x)^2}{(x+y)(2-x-y)\over 
\sqrt{x(1-x)y(1-y)}}\xi_N(y)\nn\\
&+&{\lambda^2\over 8 \pi}\int_0^1{dy\over 
\sqrt{x(1-x)y(1-y)}}\xi_N(y)
.\end{eqnarray}
This is the bosonic analog of the `t Hooft equation \cite{shei,bardeen,
halpern,tomaras}.
The equations  for $Q,\bar Q$  
are given in the next section in a slightly more general context, so 
we will not repeat it here. 
If we again use an ansatz for the $Q(u,v)$ given by 
$Q(u,v;x^-)=c_Q(x)e^{iP_-x^-}$ and the same interpretation of the symbols,
and an invariant mass, $\Lambda_Q^2=2P_-(u-v)$, we get 
\begin{eqnarray}
    \Lambda_Q^2c_Q(x)&=&\Big( {M_F^2\over x}+{M_B^2\over 1-x}\Big)c_Q(x)
-{g^2\over 2\pi}\int_0^1{dy\over (y-x)^2}{2-x-y\over \sqrt{(1-x)(1-y)}}
c_Q(y)\nn\\
&+&{\mu^2\over 4\pi}\int_0^1{dy\over \sqrt{(1-y)(1-x)}}c_Q(y)
.\end{eqnarray}
Setting $\mu^2=0$  we recover   
the equations found by Aoki\cite{aoki1, aoki2}. 
Similarly for the complex conjugate variable $\bar Q$,
 we get,
\begin{eqnarray}
    \Lambda_{\bar Q}^2c_{\bar Q}(x)&=&
\Big( {M_B^2\over x}+{M_F^2\over 1-x}\Big)c_{\bar Q}(x)
-{g^2\over 2\pi}\int_0^1{dy\over (y-x)^2}{x+y\over \sqrt{xy}}
c_{\bar Q}(y)\nn\\
&+&{\mu^2\over 4\pi}\int_0^1{dy\over \sqrt{yx}}c_{\bar Q}(y)
.\end{eqnarray}
We remark that the equation for $c_{\bar Q}$ can be  
obtained from the equation for $c_Q$ if we make the change of variable 
$x\mapsto 1-x$, and interchange  $M_B$ and $M_F$ and  use 
the principal value prescription 
(this ultimately comes from the charge conjugation invariance).

The properties of these equation have been discussed in the literature.
The two kernels above differ from the ones given in \cite{thooft1, shei,
aoki1, aoki2} by a relatively compact perturbation so they behave in the 
same way.
What is remarkable about them is that they only allow for discrete
spectrum, they do not  have  scattering states.
The corresponding eigenvectors form 
a  basis.

We make a short digression and note an interesting limit: in 
the second model let us set $g^2= 0$.
There is no coupling to gauge fields thus there is no reason 
to assume that the observables of the theory are color 
invariant. We can study this case along the same lines assuming {\it it
is a sort of mean-field approximation only} and we search for
bound states of a fermion and a boson in the linear approximation.
The Hamiltonian is quite simple,
\beq
H={1\over 4} m_B^2\int {[dp]\over |p|}N(p,p)+
{\mu^2\over 16 }\int{[dpdqdsdt]\over \sqrt{|qs|}}{ \delta[p-q+s-t]
\over t-s}Q(p,q)\bar Q(s,t)
.\eeq
The linearization is the  same as before, for the 
bound state solution   we obtain  equation (44)
with  $g^2=0$. Unfortunately this equation will not have a solution for the 
bound state energy. We need the opposite sign in the 
Hamiltonian for the coupling of $Q\bar Q$.
It is an amusing exercise to check that 
the seemingly different interaction $i\mu (\phi^{*\alpha}\psi^*_R\psi_{L\alpha}
-\psi^{*\alpha}_L\psi_R\phi_\alpha)$ produces  the same Hamiltonian, so we 
will still not find a bound state for fermion-boson pair.
We hope to  come back to some of these issues in a separate work.

\section{Beyond the linear approximation}

In this section we will discuss the equations of motion of our theory 
 in a semi-linear approximation. 
The exact equations of motion can of course be written, but it is hard to  
grasp their meaning at this point for the most general case. 
It will be interesting to look at other  approximations to see
what new information they contain.

Our first semi-linear approach is this: 
We will  keep  everything linear  in the variables  
$M$ and $N$ and terms  second order in $Q$  and $Q^\dagger$ only. 
We will  drop terms of the form $MQ, \ NQ$ and $M^2, N^2$.
Even though we have not found a justification for  why this should be a good 
approximation, we expect that  it may give us a better feeling for the 
system.
We  first   show  that this is a 
consistent approximation, that is, if the equations of motion
are also kept to the same approximation, the truncated constraints are 
preserved.
 
The constraints in this new approximation  become 
\bea 
  &\ &  M\epsilon+\epsilon M +Q\epsilon Q^\dagger=0 \nn\cr 
  &\ &  Q\epsilon + \epsilon Q=0\nn\cr 
  &\ &  \epsilon N \epsilon +N+\epsilon Q^\dagger Q=0. 
\eea 
We should also obtain semi-linearized equations of motion for these  
variables. We now show  that the linearized constraints  
are left invariant by the semi-linearized equations of motion. 
We will present the proof for a general quadratic  Hamiltonian. 
The solution of the constraint on $Q$ is simple: 
 $Q(u,v)=0$ when $u$ and $v$ have the same sign. 
We notice that the first constraint does not impose anything on 
$M(u,v)$ for $u>0,v<0$ or $u<0,v>0$, and the  
constraint is consistent since for this case we have 
$\int Q(u,q)[-\sgn(q)]\bar  
Q(q,v)[dq]=0$. 
Thus we should look at $u>0,v>0$ or both negative case for $M$
in the constraint: 
\beq 
  - 2M(u,v)+\int Q(u,q)[-\sgn(q)]\bar Q(q,v)[dq]=-2M(u,v)+\int_{-\infty}^0 [dq]
Q(u,q)\bar Q(q,v)=0 
.\eeq 
 Let us check that it is preserved by the linearized equations of motion. 
\bea 
   {\partial M(u,v) \over \partial x^-}&=&\big \{M(u,v), H\big\}\nn\cr 
    &=& 2i(h_F(u)-h_F(v))M(u,v)+
   \int[dpdrdsdt] G_1(p,r,s,t)\{M(u,v), M(p,r)M(s,t) \}\nn\cr
&+&\int [dpdrdrdt]G_3(p,r,s,t) 
 \{ M(u,v), Q(p,r)\bar Q(s,t)\})\nn\cr 
 &=&2i(h_F(u)-h_F(v))M(u,v)\nn\cr
  &+&4i\int [dpdr]G_1(p,r;v,u)M(p,r)[\sgn(u)-\sgn(v)]\nn\cr 
&-&2i\int [drdsdt]\ G_3(v,r,s,t)Q(u,r)\bar Q(s,t)+2i\int [dpdrds]\   
G_3(p,r,s,u)Q(p,r)\bar Q(s,v).  
\eea 
The equations of motion for $Q$ in this approximation becomes, 
\beq 
 {\partial Q(u,q) \over \partial x^-}=
2ih_F(u)Q(u,q)-2i\sgn(q)h_B(q)Q(u,q)+2i\int   G_3(p,r,q,u)
Q(p,r)[1-\sgn(u)\sgn(q)].  
\eeq 
Similarly for $Q^\dag$, 
\beq  
 {\partial \bar Q(q,v) \over \partial x^-}=
-2ih_F(v)\bar Q(q,v)+2i\sgn(q)h_B(q)\bar Q(q,v)-2i\int  
G_3(v,q;s,t)\bar Q(s,t)[1-\sgn(v)\sgn(q)]. 
\eeq 
Combining  these equations and using the constraint again we obtain,  
\beq 
2{\partial M(u,v)\over \partial x^-}-
\int_{-\infty}^0[{\partial Q(u,q)\over \partial  
x^-} \bar Q(q,v) +Q(u,q){\partial \bar Q(q,v) \over \partial x^-}][dq]=0 
.\eeq 
Using the same equations, we can check that 
the  condition $Q(u,v)=0$ when  $u,v$ have the same sign, 
is also preserved by the equations of motion, hence also for  
$\bar Q(u,v)$. 

We  write down the equation of motion for $N(u,v)$; 
\bea  
  {\partial N(u,v)\over \partial x^-}&=&2i[h_B(u)\sgn(u)-h_B(v)\sgn(v)] 
N(u,v)\nn\cr 
&-&4i\int [dpdq]\ G_2(p,q;v,u)[\sgn(u)-\sgn(v)]N(p,q)\nn\cr 
&+&2i\int [dpdqdt] [ G_3(p,u;q,t)Q(p,v)\bar Q(q,t)\sgn(u) 
- G_3(p,q;v,t) Q(p,q)\bar Q(u,t)\sgn(v)]. 
\eea 
Using the above equations of motion we can check 
 that the truncated constraint on $N$ is  
preserved under time evolution: 
\beq 
   (1+\sgn(u)\sgn(v)){\partial N(u,v)\over \partial x^-}-\sgn(u)
\int_{-\infty}^0 [dq]
[{\partial \bar   Q(u,q)\over \partial x^-}Q(q,v)+\bar Q(u,q){\partial
Q(q,v)\over \partial   x^-}]=0. 
\eeq 
 
Next we  discuss  the equations of motion for the unconstraint   
components. From the above equations we see that the equations for  
$Q$ and $Q^\dag$ are independent of $M$ and $N$, therefore  they can be  
solved independently. 
Furthermore, the solution acts as a source term for the  
$M$ and $N$ equations. 
Let us write down the equation of motion for $Q$ in the case of 
$u>0$ and $v<0$ for our model: 
\bea 
 {\partial Q(u,v) \over \partial x^-}&=&2ih_F(u)Q(u,v)
+2ih_B(v)Q(u,v)+i{\mu^2\over 8(u-v)}  
\int_{-{u-v\over 2}}^{{u-v \over 2}} 
[dq] {Q(q-{u-v\over 2},q+{u-v\over 2})\over \sqrt{|q+{u-v\over 2}||v|}}\nn\cr 
 &-& i{g^2\over 2}\int_{-{u-v \over 2}}^{{u-v\over 2}}[dp] 
{ p-{u\over 2}+{3v\over 2} \over (p-{u+v\over 2})^2}
{Q(p+{u-v\over 2},p-{u-v \over 2})   \over  \sqrt{|p-{u-v\over 2}||v|}}. 
\eea 
A similar equation for $\bar Q(u,v)$ holds
(which can  also be found by complex conjugation of
the $Q(v,u)$).

Notice that the equations of motion for $M(u,v)$ (for $u>0,v<0$) becomes, 
\bea  
    {\partial M(u,v)\over \partial x^-}&=& 
    \!\!\!2i(h_F(u)-h_F(v))M(u,v)
   -ig^2\int [ds] 
     {M(s+(u-v)/2,s-(u-v)/2)\over [s-(u+v)/2]^2}\nn\cr 
&+&{ig^2\over 4}\int [dqds]
{q+s\over (q-s)^2}{1 \over \sqrt{|qs|}}[Q(q+u-s,q)\bar  
 Q(s,v)- Q(u,q)\bar Q(s,v+s-q)]\nn\cr 
 &-&{i\mu^2\over 8 }\int {[dq ds]\over \sqrt{|qs|}}
\Big[{Q(u,q)\bar Q(s,s-v+q)\over  
v-q}-{Q(u+q-s,q)\bar Q(s,v)\over u-s}\Big] 
.\eea 
We note that in the above integral over $M$ we should separate the  
constrained variables from the unconstrained ones. 
At the same time we do some shift of integration variables, 
and obtain, 
\bea 
    {\partial M(u,v;x^-)\over \partial x^-}&=& i{M_F^2\over 2}[{1\over u}-
{1\over v}]M(u,v)
    \!\!\!-i  g^2\int_{-{u-v\over 2}}^{{u-v\over 2}} [ds] 
     {M(s+(u-v)/2,s-(u-v)/2)\over [s-(u+v)/ 2]^2}\nn\cr 
&+&f_+(u,v;x^-)+f_-(u,v;x^-)+g_+(u,v;x^-)+g_-(u,v;x^-)\nn\cr 
&+&Y_+(u,v;x^-)+Y_-(u,v;x^-) 
,\eea 
where all the forcing terms are functions of $Q,\bar Q$ and their 
explicit expressions are given in the appendix B. 
Note  that  once we know the solution for  
$Q$ and $Q^\dag$, $f$'s, $g$'s and $Y$'s just become time dependent  
sources for the $M$ and $N$ equations. 
Therefore we can think of this as a forced linear equation. 
Let us also write down the resulting equation of motion for $N(u,v)$, 
for $u>0,v<0$. 
\bea 
{\partial N(u,v;x^-)\over \partial x^-}&=& 
    \!\!\!i{M_B^2\over 2}[{1\over u}-{1\over v}]N(u,v)\nn\cr 
&-&i \int_{-{u-v\over 2}}^{{u-v\over 2}} [ds] 
     {N(s+(u-v)/2,s-(u-v)/2)\over  
\sqrt{|s-{u-v\over 2}||s+{u-v\over 2}| 
|uv|}}[{g^2\over 4}{(s+{3u\over 2}-{v\over 2}) 
(s+{3v\over2}-{u\over 2})\over [s-{(u+v)\over 2}]^2}-{\lambda^2\over 8}]\nn\cr 
&+&\tilde f_+(u,v;x^-)+\tilde f_-(u,v;x^-)+\tilde g_+(u,v;x^-)+
\tilde g_-(u,v;x^-)\nn\cr 
&+&\tilde Y_+(u,v;x^-)+\tilde Y_-(u,v;x^-) 
,\eea 
where we have again the forcing terms determined by the variables
  $Q, \bar Q$ 
(the explicit
formulae are given in the appendix B). 

We can give a rough argument how these equations behave.
If we look at the formulae given in the appendix B, we notice that 
the singular looking kernels are actually harmless, since the
integration regions are outside of the singular points.
This means that  once we have the solutions for the 
$Q, \bar Q$ variables
we can treat them as small perturbations to the 
equations. If we could find the Green's function for these 
linear operator equations given the sources we should be able to 
solve them. Let us assume that we have the linear equation 
$i{\partial M\over \partial x^-}={\bf L}M+S(x^-)$,
where ${\bf L}$ is a linear Hermitian operator.
If we know the eigenvectors   ${\bf L}M_\lambda=\lambda M_\lambda $
then we can use a general ansatz as
$M=\sum_\lambda a_\lambda(x^-)M_\lambda(x^-)$, and get
$a_\lambda(x^-)=-i\int_0^{x^-}dx^-<M_\lambda(x^-),S(x^-)>$.
 (In our case 
the leading singular integral operators are hermitian and 
have only discrete spectrum, hence the expansion makes sense).
This is the full solution and represents transition probabilities 
among the stationary states of the operator ${\bf L}$.
Perhaps it is better to think of the ordinary forced harmonic oscillator 
problem. When we have a time dependent forcing, this causes 
transitions between the stationary levels of the oscillator.
So, without actually solving the above equation we see that 
the forcing terms will cause transition between the stationary levels.
That physically means that the energy levels of the mesons
will have a broadening due to possible transitions.

There is another possible approximation, for which we drop all $MM$, $NN$, and 
$Q\bar Q$  terms and allow for the cross terms  
$MQ$, $NQ$ etc, and neglect any higher orders. 
In some sense this is the complementary approximation to the previous one.  
This implies that we should write the constraint as; 
\bea 
   &\ & M\epsilon+\epsilon M=0\nn\cr 
   &\ & MQ+Q\epsilon N+\epsilon Q+Q\epsilon=0\nn\cr 
   &\ & \epsilon N\epsilon +N=0 
\eea 
The first and the last one are familiar conditions. 
The middle one has the following solution (in the given approximation): 
 For $u,v>0$ (recall that $\epsilon(p)=-\sgn(p)$), 
\beq 
-2Q(u,v)+\int_{-\infty}^0 [dq] M(u,q)Q(q,v)+\int_{-\infty}^0 [dq] Q(u,q)(-\sgn(q))N(q,v)=0 
.\eeq 
For $u>0,v<0$ we have, 
\beq 
 \int_{-\infty}^0 [dq] M(u,q)Q(q,v)-\int^{\infty}_0 [dq] Q(u,q)N(q,v)=0 
.\eeq 
We satisfy the lower equation by noting that the same momenta case for  
$Q$ is given by the first constraint and the integrands then become  of 
lower order in this case.   
The consistency of these approximations could be checked. 
In fact if we write down the time derivative of the above constraint, 
\bea 
-2{\partial Q(u,v)\over \partial x^-}&+&\int_{-\infty}^0 [dq]  
({\partial M(u,q)\over \partial x^-}Q(q,v)+M(u,q){\partial Q(q,v)\over \partial x^-})\nn\cr 
&+&\int_{-\infty}^0 [dq] ({\partial Q(u,q)\over \partial x^-}N(q,v)+ 
Q(u,q){\partial N(q,v)\over \partial x^-})=0 
\eea 
To see this we use,  
\bea 
   {\partial Q(u,v;x^-)\over \partial x^-}&=& 
4i\int [dpdsdt] G_1(p,u;s,t)Q(p,v)M(s,t)\nn\cr 
&+&4i\int[dqdsdt] G_2(v,q;s,t)Q(u,q)N(s,t)\sgn(v)\nn\cr 
&+&2i(h_F(u)-\sgn (v) h_B(v)) Q(u,v) 
+2i\int [dp dq dt] G_3(p,q;v,t) Q(p,q)M(u,t)\sgn(v)\nn\cr 
&+&2i\int [dp dq ds] G_3(p,q;s,u)Q(p,q)N(s,v)\nn\cr 
&+&2i \int [dp dq] G_3(p,q;v,u)Q(p,q)[1-\sgn(u)\sgn(v)] 
.\eea
For the first time derivative in the constraint we insert this 
expression, for the time derivatives of $Q$ 
 inside the integral we only retain the 
linear terms in $Q$, since other combinations are of lower order by  
assumption. 
We should also use the  equations of motion of $M$ and $N$ for 
the opposite momenta case and only within  the  
linear approximation  as is given 
in the previous semi-linear case, we do not repeat them, higher order terms 
get multiplied by $Q$ and become small.
Then we see that the constraint is preserved within the given 
approximation.

This time we have decoupled linear equations for 
$M$ and $N$ for the opposite momenta case, since we 
ignore $Q\bar Q$ type terms, and in principle 
they can be solved independently. 
When we look at the equations for  $Q$, we should again be 
careful. The opposite momenta case are to be 
treated as independent dynamical variables: 
if  we use the  constraint equation, 
we may express the same sign momenta in terms of the solutions of
$M$ and $N$ and the opposite momenta terms of $Q$.
When we look at the opposite sign momenta 
equation for $Q$ we may separate the same sign  momenta contributions in the 
integral operators. 
But these same momenta  terms in the integral equation  
become  of higher order, since all these terms are 
multiplied by other variables, and the central part vanishes in this case,
 hence  they can be dropped. 
Let us denote the resulting  integral equation which only acts on the  
opposite momenta terms by ${\bf K}$, this is the expression 
we have found before, 
and write the remaining parts as an abstract integral operator 
${\bf F}(x^-)$. Notice that it has  dependence on $x^-$ via the solutions of 
$M$ and  $N$. 
The time dependence of $M$ and  $N$ are  rather simple for this case,
since we have singular integral equations with discrete spectra. We can 
in principle substitute the solutions we picked  into this  equation. 
Hence we have an integral equation  
\beq  
{\partial Q(u,v;x^-)\over \partial x^-}=[{\bf K} Q](u,v;x^-)+
[{\bf F}(x^-)Q](u,v;x^-) 
.\eeq
It is most natural to think of the last term as a time dependent 
perturbation. 
We can write this  perturbation term ${\bf F}(x^-)$: 
\bea 
  [{\bf F}(x^-)Q](u,v;x^-)&=&-{ig^2\over 2}\Big[\int_0^u [ds] \int_0^{u-s}[dp]+
\int_{-\infty}^0 [ds] 
 \int_{u-s}^\infty [dp]\Big] {M(s,s+p-u;x^-)\over (s-u)^2}Q(p,v)\nn\cr 
&+&i\Big[\int_{-\infty}^v [dq] \int_0^{v-q} [dt]+
\int_v^0[dq]\int_{v-q}^0[dt]\Big] 
\Big[ {g^2\over 8}{(t+q)^2+(q-v)^2-2q^2\over (v-t)^2} 
-{\lambda^2\over 16}\Big]\nn\cr
&\times& {N(t+q-v,t;x^-)\over \sqrt{|vq(t+q-v)|}}Q(u,q)\nn\cr 
&-&i\Big[\int_v^0 [dt] \int_0^{t-v} [dp] +
\int_{-\infty}^v [dt] \int_{t-v}^0 [dp]\Big] 
\Big({\mu^2\over 8(t-v)}+{g^2\over 4}{p-t+2v\over (p-t)^2}\Big)\nn\cr 
&\times& {M(u,t;x^-) \over  
\sqrt{|(p-t+v)v|}}Q(p,p-t+v)\nn\cr 
&+&i\Big[\int_0^u [ds] \int_0^{u-s} [dp] +
\int_u^{\infty} [ds] \int_{u-s}^0 [dp]\Big] 
\Big( {\mu^2\over 8 (u-s)}+{g^2\over 4}{p+2s-u\over (p-u)^2}\Big)\nn\cr 
&\times& {N(s,v;x^-)\over \sqrt{|(p+s-u)s|}}Q(p,p+s-u) 
\eea 
 
The method of solving such equations is known in principle.
We can treat the last term as a truly time dependent perturbation, 
but 
this time it involves the unknown itself and thus cannot 
be  solved in closed form. However, 
we can solve it perturbatively. The kernels again do not become 
singular within the given ranges of the integrals except at the 
boundaries. The singularities are not as severe and   
we expect that the  perturbations are  small,   
so that one can obtain a reasonable answer from this approach.
We will not go into further  details, but the basic result is 
again the possibility of transitions between the different levels
of the boson-fermion mesons due to the interactions.

\section{Acknowledgements}
First of all we would like to thank S. G. 
Rajeev for various useful discussions and 
suggestions. 
E. Langmann's efforts   improved the presentation 
considerably.  The revision is done  while O.T. Turgut 
is in KTH, as a Gustafsson fellow. 
We thank   K. Bardakci for useful discussions. A. Konechny  thanks
Feza Gursey Institute for hospitality  during the summer
of 99 where this work has began. O. T. Turgut would like to  acknowledge
discussions with M. Arik, K. Gawedzki, J. Gracia-Bondia, 
J. Mickelsson, R. Nest, Y. Nutku, C. Saclioglu, M. Walze.
He also would like to thank E. Schrohe for an opportunity to present this work
in Potsdam and for the hospitality, as well as to  W. Zakrzewski, for the kind
invitation to talk in Durham and for the  hospitality of  the theory group,
and Lawrence Berkeley National Lab. for  the invitation to complete this work.
The work of A.~K. was partially supported by the Director, Office of Energy 
Research, Office of High Energy and Nuclear Physics, Division of High 
Energy Physics of the U.S. Department of Energy under Contract 
DE-AC03-76SF00098 and  partially by the National Science Foundation grant PHY-95-14797

\section{Appendix A: Baryon number}

We define the  Fredholm operators in a ${\bf Z}_2$ graded  context
as in the ordinary case (there is an   extension to 
the ${\bf Z}$ grading which should fit  
to our model better: The definition of the Fredholm operator
shows that the body is an ordinary Fredholm operator and 
the rest is compact. So below use the body for all the formulae, and 
take  the super-trace of $\Phi-\hat \epsilon$'s body part):
 a Fredholm operator is an  invertible operator up to compact
operators. This  again implies the kernel and the cokernel are 
actually finite dimensional.
Let us write down the kernel in a decomposition $V_e|V_o$,
and define a super-dimension, which is the dimension of the even
part of the kernel minus the dimension of the odd part.
Sdim$({\rm Ker}(A))={\rm dim }(P_e{\rm Ker}(A)P_e)-{\rm dim}(P_o{\rm
  Ker}(A)P_o)$,
where $P_e$ and $P_o$ denote projections onto the even and odd 
subspaces respectively.
Then the index should  be, 
\beq
      {\rm SInd} (A)={ \rm Sdim} ({\rm Ker }(A))-{\rm Sdim}({\rm Coker}(A))
.\eeq
We can extend  
 the Calderon  theorem to our   
case (see \cite{noncomm} for a good introduction and the original result):
If we have an operator $A$ which is Fredholm,
and assume we have an operator $B$ such that 
$(I-AB)^m$ and $(I-BA)^m$ are trace class for an integer $m$,
then we can compute the super-Fredholm index as
$\Str(I-BA)^m-\Str(I-AB)^m$.
Let us now see that in our problem the supertrace of $\Phi$ is 
indeed this index.
It will be more convenient to decompose our operator with respect to 
the positive and negative subspaces,
thus we write everything with respect to
${\cal H}_+|{\cal H}_+\oplus{\cal H}_-|{\cal H}_-$, we do not repeat 
odd and even superscripts, 
since the bar indicates this separation.
In this decomposition our group conditions can be 
found from,
\beq
     g^\dag E g=E \quad g Eg^\dag=E\quad E=\pmatrix{\epsilon &0\cr 0& 1}
,\eeq
so $E$ is the same as before in this matrix representation, 
it is  interpreted differently.
The orbit is  with respect to this decomposition,
\beq
    \Phi=g\hat \epsilon g^{-1} \quad \hat \epsilon=\pmatrix{-1&0\cr 0& 1}
.\eeq
So if we write  
$g:{\cal H}_+|{\cal H}_+\oplus{\cal H}_-|{\cal H}_- \to
{\cal H}_+|{\cal H}_+\oplus{\cal H}_-|{\cal H}_-$,
 explicitly, we have 
\beq
  g=\pmatrix{A&B\cr C&D} \quad B,\ C\in {\cal I}_2({\cal H}_\mp|{\cal H}_\mp,
{\cal H}_\pm|{\cal H}_\pm)
.\eeq
From the first group condition  we get,
$A^\dag\epsilon  A+C^\dag  C=\epsilon$ and
$D^\dag  D + B^\dag\epsilon  B= 1$
and from the second one we get,
$A\epsilon A^\dag + B B^\dag =\epsilon$ and 
$D D^\dag +C\epsilon C^\dag=1$.
Since $B,C$ are Hilbert-Schmidt in the more generalized sense,
we have $A,D$ super Fredholm.
Further we can use the above theorem to compute the 
index of $A,D$, for example 
\beq
   {\rm SInd}(D)=\Str(B^\dag \epsilon B)-\Str(C\epsilon C^\dag)
.\eeq
Let us compute the conditional supertrace of $\Phi-\hat \epsilon$,
(in fact we see that this is the correct way we should be 
computing it), first we write it explicitly with 
respect to the above decomposition,
\beq 
 g\hat\epsilon g^{-1}-\hat \epsilon=\pmatrix{-A\epsilon A^\dag \epsilon
+B B^\dag \epsilon+1& *\cr
*& -C\epsilon  C^\dag+D \epsilon D^\dag-1}
.\eeq
If we use the above group properties, we get 
\beq
    \Phi-\hat \epsilon=\pmatrix{2BB^\dag\epsilon &*\cr *&-2C\epsilon C^\dag }
.\eeq
The conditional supertrace of this operator  gives us,
$\Str_\epsilon(\Phi-\hat \epsilon)=2(\Str(BB^\dag\epsilon )-
\Str(C \epsilon C^\dag))$, which is equal to  $2$SInd$(D)$
(using $\Str(BB^\dag\epsilon)=\Str(B^\dag\epsilon B)$).
Thus we prove using only the geometry of the 
super-Grassmanian that this is an integer.

\section{Appendix B: Forcing terms} 

Here we present the forcing functions for the inhomogeneous equations of 
the  previous section. The ones we got for $M$ in the first semi-linear approximation 
is given by,

\bea 
    f_+(u,v;x^-)&=&-i{g^2\over 2}\int_{{u-v \over 2}}^{\infty}[dp]
\int_{-\infty}^{0}[dq] 
{ Q(p-{u-v\over 2},q)\bar Q(q,p+{u-v\over 2})\over [p-{u+v\over 2}]^2}\nn\cr 
    f_-(u,v;x^-)&=&-i{g^2\over 2}\int^{-{u-v \over 2}}_{-\infty}[dp]
\int^{\infty}_{0}[dq] 
{ Q(p-{u-v\over 2},q)\bar Q(q,p+{u-v\over 2})\over [p-{u+v\over 2}]^2}\nn\cr 
    g_+(u,v;x^-)&=&i{g^2\over 4}\Big[\int_{0}^{u}[ds] 
\int_{-{u-s\over 2}}^{{u-s\over 2}}[dq] 
+\int_{u}^{\infty}[ds]\int_{-{s-u\over 2}}^{{s-u\over 2}}[dq]\Big] 
 {q+{3s\over 2}-{u\over 2}\over [q-{u+s\over 2}]^2} 
{Q(q+{u-s\over 2},q-{u-s\over 2})\bar Q(s,v) \over  
\sqrt{|q-{u-s\over 2}||s|}}\nn\cr 
    g_-(u,v;x^-)&=&-i{g^2\over 4}\Big[\int_{v}^{0}[ds] 
\int_{-{s-v\over 2}}^{{s-v\over 2}}[dq] 
+\int^{v}_{-\infty}[ds]\int_{-{v-s\over 2}}^{{v-s\over 2}}[dq]\Big] 
 {q+{3s\over 2}-{v\over 2}\over [q-{v+s\over 2}]^2} 
{Q(u,s)\bar Q(q-{v-s\over 2},q+{v-s\over 2}) \over  
\sqrt{|q-{v-s\over 2}||s|}} 
\eea 
\bea 
   Y_-(u,v;x^-)&=&+{i\mu^2\over 8}\Big[\int_{0}^{u}[ds] 
\int_{-{u-s\over 2}}^{{u-s\over 2}}[dq] 
+\int_{u}^{\infty}[ds]\int_{-{s-u\over 2}}^{{s-u\over 2}}[dq]\Big] 
{Q(q+{u-s\over 2},q-{u-s\over 2})\bar Q(s,v) \over (u-s) 
\sqrt{|q-{u-s\over 2}||s|}}\nn\cr 
 Y_+(u,v;x^-)&=&-{i\mu^2\over 8}\Big[\int_{v}^{0}[ds] 
\int_{-{s-v\over 2}}^{{s-v\over 2}}[dq] 
+\int^{v}_{-\infty}[ds]\int_{-{v-s\over 2}}^{{v-s\over 2}}[dq]\Big] 
{Q(u,s)\bar Q(q-{v-s\over 2},q+{v-s\over 2}) \over (v-s)  
\sqrt{|q-{v-s\over 2}||s|}} 
.\eea 
 
The forcing terms for the first semi-linear approximation
for the $N$ variable,
\bea 
    \tilde f_-(u,v;x^-)&=&-i\int_{{u-v \over 2}}^{\infty}[dp]
\int_{-\infty}^{0}[dq] 
{ \bar Q(p+{u-v\over 2},q) Q(q,p-{u-v\over 2})\over \sqrt{|p-{u-v\over 2}|| 
p+{u-v\over 2}||uv|}} 
\Big[{g^2\over 8}{(p+{3u\over 2}-{v\over 2})
(p+{3v\over 2}-{u\over 2})\over [p-{u+v\over 2}]^2} 
 -{\lambda^2\over 4}\Big]\nn\cr 
    \tilde f_+(u,v;x^-)&=&i\int^{-{u-v \over 2}}_{-\infty}[dp]
\int^{\infty}_{0}[dq] 
{ \bar Q(p+{u-v\over 2},q) Q(q,p-{u-v\over 2})\over  \sqrt{|p-{u-v\over 2}|| 
p+{u-v\over 2}||uv|}} 
\Big[{g^2\over 8}{(p+{3u\over 2}-{v\over 2})(p+{3v\over 2}-{u\over 2})  
\over [p-{u+v\over 2}]^2}-{\lambda^2\over 4}\Big]\nn\cr 
    \tilde g_+(u,v;x^-)&=&i{g^2\over 4}\Big[\int_{0}^{u}[dp] 
\int_{-{u-p\over 2}}^{{u-p\over 2}}[ds] 
+\int_{u}^{\infty}[dp]\int_{-{p-u\over 2}}^{{p-u\over 2}}[ds]\Big] 
 {s-{p\over 2}+{3u\over 2}\over [s-{u+p\over 2}]^2} 
{Q(p,v) Q(s+{u-p\over 2},s-{u-p\over 2}) \over  
\sqrt{|s-{p-u\over 2}||u|}}\nn\cr 
    \tilde g_-(u,v;x^-)&=&i{g^2\over 4}\Big[\int_{v}^{0}[dp] 
\int_{-{p-v\over 2}}^{{p-v\over 2}}[ds] 
+\int^{v}_{-\infty}[dp]\int_{-{v-p\over 2}}^{{v-p\over 2}}[ds]\Big] 
 {q+{3s\over 2}-{v\over 2}\over [q-{v+s\over 2}]^2} 
{ Q(s-{v-p\over 2},s+{v-p\over 2}) \bar Q(u,p)\over  
\sqrt{|s-{p-v\over 2}||v|}} 
\eea 
\bea 
   \tilde Y_+(u,v;x^-)&=&-{i\mu^2\over 8}\Big[\int_{0}^{u}[dp] 
\int_{-{u-p\over 2}}^{{u-p\over 2}}[ds] 
+\int_{u}^{\infty}[dp]\int_{-{p-u\over 2}}^{{p-u\over 2}}[ds]\Big] 
{Q(p,v)\bar Q(s+{u-p\over 2},s-{u-p\over 2}) \over (p-u) 
\sqrt{|s+{u-p\over 2}||u|}}\nn\cr 
 \tilde Y_-(u,v;x^-)&=& {i\mu^2\over 8}\Big[\int_{v}^{0}[dp] 
\int_{-{p-v\over 2}}^{{p-v\over 2}}[ds] 
+\int^{v}_{-\infty}[dp]\int_{-{v-p\over 2}}^{{v-p\over 2}}[ds]\Big] 
{Q(s-{v-p\over 2},s+{v-p\over 2})\bar Q(u,p) \over (p-v)  
\sqrt{|s+{p-v\over 2}||v|}} 
.\eea


\begin{thebibliography}{50}

\bibitem{thooft2} G. `t Hooft, Nuc. Phys. {\bf B 75} (1974) 461. 
\bibitem{coote} C. G. Callan,  N. Coote, and J. D. Gross, 
   Phys. Rev. {\bf D 13} (1976) 1649.
\bibitem{einhorn} M. B. Einhorn, Phys. Rev {\bf D 14}(1976) 3451,
Phys. Rev. {\bf D 15} (1976) 3037.
\bibitem{pak} N. K. Pak and H. C. Tse, Phys. Rev. {\bf D 14} (1976) 3472.
\bibitem{bars} I. Bars and M. B. Green, Phys. Rev. {\bf D 17} (1978) 537.

\bibitem{istlect} S. G. Rajeev, {\it Derivations of hadronic 
structure functions from QCD},{\it Conformal Field Theory} edited by
Y. Nutku, C. Saclioglu, and O. T. Turgut, Perseus Books, 2000. 

\bibitem{2dqhd} S. G. Rajeev, Int. J. Mod. Phys.{\bf A 9} 5583 (1994).

\bibitem{wadia}  A. Dhar, G. Mandal, and  
S. R.  Wadia, Nuc. Phys. {\bf B 436} (1994) 487. 
\bibitem{mandal} A. Dhar, P. Lakdawala, G. Mandal, and S. Wadia,  
Intr. Jour. Mod. Phys. {\bf A 10} (1995) 2189.   

\bibitem{thooft1} G. `t Hooft, Nuc. Phys. {\bf B 72} (1974) 461.  
\bibitem{witten1} E. Witten, Nucl. Phys. {\bf B 160} (1979) 57.

\bibitem{2dbaryon} P. Bedaque, I. Horvath and S. G. Rajeev,  
Mod. Phys. Lett.  {\bf A7} (1992) 3347. 

\bibitem{halpern3} M. B. Halpern and C. Schwartz, 
Int. J. Mod. Phys. {\bf A 14} (1999) 3059,
Int. J. Mod. Phys. {\bf A 14} (1999) 4653.

\bibitem{lee} S. G. Rajeev and C. W-H. Lee, Int. J. Mod. Phys. 
{\bf A 14} (1999) 4653, J. Math. Phys. {\bf 39} (1998) 5199,
Nucl. Phys. {\bf B 529} (1998) 656.

\bibitem{shei} S. S. Shei and H. S. Tsao, Nucl. Phys. {\bf B 141} (1978)
445.
\bibitem{bardeen} W. A. Bardeen and P. B. Pearson, Phys. Rev.
{\bf D 14}(1976) 547.
\bibitem{halpern} M. B. Halpern and P. Senjanovic, Phys. rev 
{\bf D 15}(1977) 1655.
\bibitem{rajteo} S. G. Rajeev and O.T.Turgut, Comm. Math. Phys. {\bf 192} 
  493 (1998). 
\bibitem{tomaras} T. N. Tomaras, Nuc.  Phys. {\bf  B 163} (1980) 79. 



\bibitem{aoki1} K. Aoki, Phys. Rev. {\bf D 49} (1994) 573.
\bibitem{aoki2} K. Aoki and T. Ichihara, Phys. Rev. {\bf D 52} (1995) 6435.
\bibitem{cavicchi} M. Cavicchi, Intr. Jour. Mod. Phys. {\bf A 10}  
(1995) 167. 

\bibitem{pinsky} O. Lunin and S. Pinsky, Phys. Rev. {\bf D 63} (2001) 45019.

\bibitem{jackiw} R. Jackiw and Rajaraman, Phys. Rev. Let.
{\bf 54}(1985) 1219, Erratum-ibid {\bf 54}(1985) 2060.  

\bibitem{langmann1} H. Grosse and E. Langmann, Jour. Math. Phys.
{\bf 33} (1992) 1032.

\bibitem{langmann2} H. Grosse and E. Langmann,  Lett. Math. Phys.
{\bf 21} (1991) 69.

\bibitem{langmann3} E. Langmann, Jour. Math. Phys. {\bf 35} (1994)
96.  

\bibitem{gohberg} I. C. Gohberg and M. G. Krein,{\it  
Introduction to The Theory of Nonself-adjoint Linear Operators}, Trans. 
Am. math. Soc. 1968. 
\bibitem{simon} B. Simon,{\it Trace Ideals and Their Applications}, 
Cambridge Univ. Press, Cambridge, 1979. 

\bibitem{superdisk} O. T. Turgut,  Jour. Math. Phys. {\bf 42} (2001) 4259.

\bibitem{noncomm} J. M. Gracia-Bondia, J. C. Varilly and 
H. Figueroa, {\it Elements of Noncommutative Geometry},
Birkhauser Boston (2000).


 
\end{thebibliography}
\end{document}